\documentclass[12pt]{article}
\usepackage{epsfig}
\usepackage{amssymb,amsmath}

\newcommand{\bea}{\begin{eqnarray}}
\newcommand{\eea}{\end{eqnarray}}

 \makeatletter
\def\alt{\mathrel{\mathpalette\gl@align<}}
\def\agt{\mathrel{\mathpalette\gl@align>}}
\def\gl@align#1#2{\lower.6ex\vbox{\baselineskip\z@skip\lineskip\z@
\ialign{$\m@th#1\hfil##\hfil$\crcr#2\crcr\sim\crcr}}} \makeatother

\begin{document}
%
\vspace*{1.0cm}

\begin{center}
\baselineskip 20pt 
{\Large\bf 
Simple brane-world inflationary models: an update
}
\vspace{1cm}

{\large 
Nobuchika Okada$^{~a}$  and  Satomi Okada$^{~b}$
}
\vspace{.5cm}

{\baselineskip 20pt \it
$^a$Department of Physics and Astronomy, University of Alabama, Tuscaloosa, AL35487, USA\\
$^b$Graduate School of Science and Engineering, Yamagata University,  \\
Yamagata 990-8560，Japan} 

\vspace{.5cm}

\vspace{1.5cm} {\bf Abstract}
\end{center}

In the light of the Planck 2015 results, 
  we update simple inflationary models based on the quadratic, quartic, Higgs and Coleman-Weinberg potentials 
  in the context of the Randall-Sundrum brane-world cosmology. 
Brane-world cosmological effect alters the inflationary predictions of the spectral index ($n_s$) and 
  the tensor-to-scalar ratio ($r$) from those obtained in the standard cosmology.
In particular, the tensor-to-scalar ratio is enhanced in the presence of the 5th dimension. 
In order to maintain the consistency with the Planck 2015 results for the inflationary predictions in the standard cosmology,   
  we find a lower bound on the five-dimensional Planck mass ($M_5$). 
On the other hand, the inflationary predictions laying outside of the Planck allowed region 
  can be pushed into the allowed region by the brane-world cosmological effect 
  with a suitable choice of $M_5$.

\thispagestyle{empty}

\newpage

\addtocounter{page}{-1}

\baselineskip 18pt

\section{Introduction} 
Inflationary universe is the standard paradigm in the modern cosmology~\cite{inflation1, inflation2, chaotic_inflation, inflation4} 
  which provides not only solutions to various problems in the standard big-bang cosmology, 
  such as the flatness and horizon problems, but also the primordial density fluctuations 
  which are necessary for the formation of  the large scale structure observed in the present universe.  
Various inflationary models have been proposed with typical inflationary predictions for the primordial perturbations. 
Recent cosmological observations by, in particular, the Wilkinson Microwave Anisotropy Probe (WMAP)~\cite{WMAP9} 
  and the Planck satellite~\cite{Planck2013} experiments have measured the cosmological parameters precisely and 
  provided constraints on the inflationary predictions for the spectral index ($n_s$), 
  the tensor-to-scalar ratio ($r$), the running of the spectral index ($\alpha=d n_s/d \ln k$), 
  and non-Gaussianity of the primordial perturbations. 
It is expected that future cosmological observations will become more precise, and allow us to discriminate inflationary models. 
Very recently, the Planck collaboration has released the Planck 2015 results~\cite{Planck2015}, 
  which provide us with the most stringent constraints on the inflationary predictions.

The brane-world cosmology is based on a 5-dimensional model first proposed 
  by Randall and Sundrum (RS) \cite{RS2}, the so-called RS II model, 
  where the standard model particles are confined on a "3-brane"  at a boundary embedded 
  in 5-dimensional anti-de Sitter (AdS) space-time. 
Although the 5th dimensional coordinate tends to infinity,  
  the physical volume of the extra-dimension is finite because of the AdS space-time geometry. 
The 4-dimensional massless graviton is localizing around the brane on which the Standard Model 
  particles reside, while the massive Kaluza-Klein gravitons are delocalized toward infinity, 
  and as a result, the 4-dimensional Einstein-Hilbert action is reproduced at low energies. 
Cosmology in the context of the RS II setup has been intensively studied~\cite{braneworld} 
  since the finding of a cosmological solution in the RS II setup~\cite{RS2solution}. 
Interestingly, the Friedmann equation in the RS brane-world cosmology leads to a non-standard 
  expansion law of our 4-dimensional universe at high energies, 
  while reproducing the standard cosmological law at low energies. 
This non-standard evolution of the early universe causes modifications of  
  a variety of phenomena in particle cosmology, such as the dark matter relic abundance~\cite{DM_BC} 
  (see also \cite{DM_GB} for the modification of dark matter physics in a more general brane-world cosmology, 
   the Gauss-Bonnet brane-world cosmology~\cite{GB}), 
   baryogensis via leptogenesis~\cite{LG_BC}, and gravitino productions in the early universe~\cite{gravitino_BC}.

The modified Friedmann equation also affects inflationary scenario. 
A chaotic inflation with a quadratic inflaton potential has been examined in \cite{inflation_BC},  
  and it has been shown that the inflationary predictions are modified from those in the 4-dimensional standard cosmology. 
It is remarkable that the power spectrum of tensor fluctuation is found to be enhanced 
  in the presence of the 5-dimensional bulk ~\cite{PT_BC}. 
Taking this brane-cosmological effect into account, simple monomial inflaton potentials 
   have been analyzed in \cite{inflation_models_BC, inflation_models_BC2} 
   in light of the WMAP 3yr results and Planck 2013 results, respectively.

Last year, the Background Imaging of Cosmic Extragalactic Polarization (BICEP2) collaboration 
   reported their observation of CMB $B$-mode polarization~\cite{BICEP2}, 
   which was interpreted as the primordial gravity waves with $r=0.20^{+0.07}_{-0.05}$ (68\% confidence level)
   generated by inflation.  
Motived by the BICEP2 result of the large tensor-to-scalar ratio of ${\cal O}(0.1)$,
   various inflationary models and their predictions have been reexamined/updated. 
See, for example, Ref.~\cite{OSS2014} for an update of the inflationary predictions of simple models 
   in the standard cosmology. 
In \cite{OO2}, we have examined simple inflationary models in the context of the RS brane-world cosmology, 
   and found that the brane-world cosmological effect enhances the tensor-to-scalar ratio to nicely 
   fit the BICEP2 result.\footnote{
See \cite{OO3} for discussion about simple inflationary models in the Gauss-Bonnet brane-world cosmology. 
} 
However, recent joint analysis of BICEP2/Keck Array and Planck data~\cite{BKP} has concluded 
  that uncertainty of dust polarization dominates the excess observed by the BICEP2 experiment. 
Now the Planck 2015 results set an upper bound on the tensor-to-scalar ratio to be $r \lesssim 0.1$. 
The purpose of this paper is to update simple brane-world inflationary models 
  discussed in \cite{OO2}, in light of the Planck 2015 results.  
The fact that the brane-world inflationary models could nicely fit the BICEP2 result 
  implies that the Planck 2015 results constrain the brane-world cosmological effect.

In the next section, we briefly review the RS brane-world cosmology and 
   present basic formulas which will be employed in our analysis for inflationary predictions. 
We first update in Sec.~3 the inflationary predictions of the textbook inflationary models 
   with the quadratic and quartic potentials.  
In Secs.~4 and 5, we analyze the Higgs potential and the Coleman-Weinberg potential models, 
   respectively, with various values of the inflaton vacuum expectation value (VEV) and 
   the 5-dimensional Planck mass in the brane-world cosmology. 
The last section is devoted to conclusions.

\section{Inflationary scenario in the brane-world cosmology}
In the RS II brane-world cosmology, the Friedmann equation for a spatially flat universe 
  is found to be~\cite{RS2solution}
\begin{equation}
H^2 = \frac{\rho}{3 M_P} \left(1+\frac{\rho}{\rho_0} \right) + \frac{C}{a^4},
\label{BraneFriedmannEq}
\end{equation}
where $M_P=2.435 \times 10^{18}$ GeV is the reduced Planck mass, 
\begin{eqnarray}
\rho_0 = 12 \frac{M_5^6}{M_P^2},
\label{rho_0}
\end{eqnarray}
with $M_5$ being the 5-dimensional Planck mass, 
 a constant $C$ is referred to as the ``dark radiation,'' 
 and we have omitted the 4-dimensional cosmological constant. 
Note that the Friedmann equation in the standard cosmology is reproduced for  $\rho/\rho_0 \ll 1$ and $\rho/(3 M_P^2) \gg C/a^4$. 
There are phenomenological, model-independent constraints for these new parameters 
  from Big Bang Nucleosynthesis (BBN), which provides successful explanations for synthesizing light nuclei in the early universe. 
In order not to ruin the success, the expansion law of the universe must obey the standard cosmological one 
  at the BBN era with a temperature of the universe $T_{\rm BBN} \simeq 1$ MeV.  
Since the constraint on the dark radiation is very severe~\cite{dark_radiation}, we simply set $C=0$ in the following analysis.
We estimate a lower bound on $\rho_0$ by $\rho_0^{1/4} \gtrsim T_{\rm BBN}$ and find $M_5 \gtrsim 8.8$ TeV. 
A more sever constraint is obtained from the precision measurements of the gravitational law in sub-millimeter range. 
Through the vanishing cosmological constant condition, we find $\rho_0^{1/4} \gtrsim 1.3$ TeV, 
  equivalently, $M_5 \gtrsim 1.1 \times 10^8$ GeV as discussed in the original paper by Randall and Sundrum~\cite{RS2}.
The energy density of the universe is high enough to satisfy $\rho/\rho_0 \gtrsim1$, 
 the expansion law becomes non-standard, and this brane-world cosmological effect alters 
 results obtained in the context of the standard cosmology.

Let us now consider inflationary scenario in the brane-world cosmology with the modified Friedmann equation. 
In slow-roll inflation, the Hubble parameter is approximately given by (from now on, we use the Planck unit, $M_P=1$)
\begin{eqnarray}
  H^2 = \frac{V}{3} \left( 1+ \frac{V}{\rho_0}\right), 
\label{H_BC}
\end{eqnarray}
  where $V(\phi)$ is a potential of the inflaton $\phi$.   
Since the inflaton is confined on the brane, the power spectrum of scalar perturbation obeys  
  the same formula as in the standard cosmology, except for the modification of the Hubble parameter~\cite{inflation_BC}, 
\begin{eqnarray}
   {\cal P}_{\cal S} =\frac{9}{4 \pi^2} \frac{H^6}{(V')^2}, 
\end{eqnarray}
  where the prime denotes the derivative with respect to the inflaton field $\phi$. 
 From the Planck 2015 results~\cite{Planck2015}, 
   we fix the power spectrum of scalar perturbation as $ {\cal P}_{\cal S}(k_0) = 2.196 \times 10^{-9}$ 
   for the pivot scale chosen at $k_0=0.002$ Mpc$^{-1}$.  
The spectral index is given by
\begin{eqnarray}
 n_s -1 = \frac{d \ln {\cal P}_{\cal S}}{d\ln k} =-6 \epsilon + 2 \eta 
\end{eqnarray}
  with the slow-roll parameters defined as  
\begin{eqnarray}
\epsilon = \frac{V^\prime}{6 H^2} \left( \ln H^2 \right)^\prime, \; \; \eta =\frac{V^{\prime \prime}}{3 H^2} .
\end{eqnarray}
The running of the spectral index, $\alpha=dn_s/d\ln k$,  is given by 
\begin{eqnarray}
   \alpha=\frac{dn_s}{d\ln k} =\frac{V^\prime}{3 H^2}  \left( 6 \epsilon^\prime - 2 \eta^\prime   \right).  
\end{eqnarray}
On the other hand, in the presence of the extra dimension where graviton resides, 
   the power spectrum of tensor perturbation is modified to be~\cite{PT_BC} 
\begin{eqnarray}
 {\cal P}_{\cal T} =8  \left(  \frac{H}{2 \pi} \right)^2 F(x_0)^2, 
 \label{PT}
 \label{P_T}
\end{eqnarray}
where  $x_0 = 2 \sqrt{3 H^2/\rho_0}$, and 
\begin{eqnarray}
 F(x)= \left( \sqrt{1+x^2} - x^2 \ln\left[ \frac{1}{x}+\sqrt{1+\frac{1}{x^2}} \right]   \right)^{-1/2}. 
\end{eqnarray} 
For $x_0 \ll 1$ (or $V/\rho_0 \ll1$), $F(x_0) \simeq 1$, and Eq.~(\ref{P_T}) reduces to 
 the formula in the standard cosmology. 
For $x_0 \gg 1$ (or $V/\rho_0 \gg1$), $F(x_0) \simeq \sqrt{3 x_0/2} \simeq \sqrt{3 V/\rho_0} \gg 1$. 
The tensor-to-scalar ratio is defined as $r =  {\cal P}_{\cal T}/ {\cal P}_{\cal S}$.

The e-folding number is given by 
\begin{eqnarray}
N_0 = \int_{\phi_e}^{\phi_0} d\phi \; \frac{3 H^2}{V^\prime} = \int_{\phi_e}^{\phi_0} d\phi \; \frac{V}{V^\prime} \left(1+\frac{V}{\rho_0} \right), 
\end{eqnarray}
where $\phi_0$ is the inflaton VEV at horizon exit of the scale corresponding to $k_0$, 
  and $\phi_e$ is the inflaton VEV at the end of inflation, which is defined by ${\rm max}[\epsilon(\phi_e), | \eta(\phi_e)| ]=1$.
In the standard cosmology, we usually consider $N_0=50-60$ in order to solve the horizon problem. 
Since the expansion rate in the brane-world cosmology is larger than the standard cosmology case, 
  we may expect a larger value of the e-folding number. 
For the model-independent lower bound,  $\rho_0^{1/4} \gtrsim 1$ MeV, the upper bound $N_0 \lesssim 75$ 
   was found in \cite{N_BC}. 
In what follows, we consider $N_0=50$, $60$, and $70$, as reference values.

\section{Textbook inflationary models}
We first analyze the textbook chaotic inflation model with a quadratic potential~\cite{chaotic_inflation},  
\begin{eqnarray}
V =\frac{1}{2} m^2 \phi^2. 
\end{eqnarray}
In the standard cosmology, simple calculations lead to the following inflationary predictions:
\begin{eqnarray}
n_s=1-\frac{4}{2 N_0+1},  \; \; 
r=\frac{16}{2 N_0+1}, \; \; 
\alpha= - \frac{8}{(2 N_0+1)^2}. 
\end{eqnarray}
The inflaton mass is determined so as to satisfy the power spectrum measured by the Planck satellite experiment, 
 ${\cal P}_{\cal S}(k_0)=2.196 \times 10^{-9}$:
\begin{eqnarray}
 m [{\rm GeV}]= 1.45 \times 10^{13}  \left(  \frac{121}{2 N_0+1}\right). 
\end{eqnarray}

In the brane-world cosmology, these inflationary predictions in the standard cosmology 
  are altered due to the modified Friedmann equation. 
In the limit, $V/\rho_0 \gg 1$,  the Hubble parameter is simplified as $H^2 \simeq  V^2/(3 \rho_0)$, 
 and we can easily find 
\begin{eqnarray}
n_s=1-\frac{5}{2 N_0+1},  \; \; 
r=\frac{24}{2 N_0+1}, \; \; 
\alpha= - \frac{10}{(2 N _0+1)^2}. 
\label{phi2_BClimit}
\end{eqnarray}
The initial ($\phi_0$) and the final ($\phi_e$) inflaton VEVs are found to be 
\begin{eqnarray}
 \phi_0^4 =96 \frac{M_5^6}{m^2} (2 N_0+1), \; \;  \phi_e^4 =96 \frac{M_5^6}{m^2} . 
 \label{phi2_int}
\end{eqnarray}
For a common $N_0$ value, the spectral index is reduced, 
 while $r$ and $|\alpha |$ are found to be larger than those predicted in the standard cosmology. 
In the brane-world cosmology, once $\rho_0$ is fixed, equivalently, $M_5$ is fixed through Eq.~(\ref{rho_0}), 
 the inflaton mass is determined by the constraint ${\cal P}_{\cal S}(k_0)=2.196 \times 10^{-9}$. 
For the limit $V/\rho_0 \gg 1$, we find~\cite{inflation_BC}
\begin{eqnarray}
 \frac{m}{M_5} \simeq 1.26 \times 10^{-4} \left(  \frac{121}{2 N_0+1}\right)^{5/6} . 
 \label{m/M5}
\end{eqnarray} 
Note that, unlike inflationary scenario in the standard cosmology, the inflaton becomes lighter as $M_5$ is lowered. 
This analysis is valid for $V(\phi_0)/\rho_0 \gg 1$, in other words, $M_5 \ll 0.01$ 
  by using Eqs.~(\ref{rho_0}), (\ref{phi2_int}), and (\ref{m/M5}).

\begin{figure}[htbp]
\begin{center}
\includegraphics[width=0.45\textwidth,angle=0,scale=1.04]{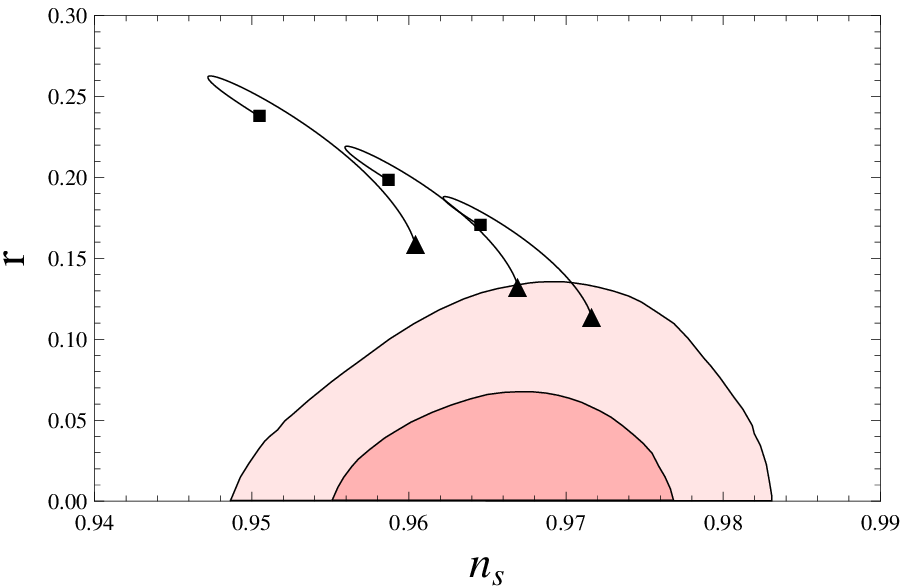} \hspace{0.5cm}
\includegraphics[width=0.45\textwidth,angle=0,scale=1.08]{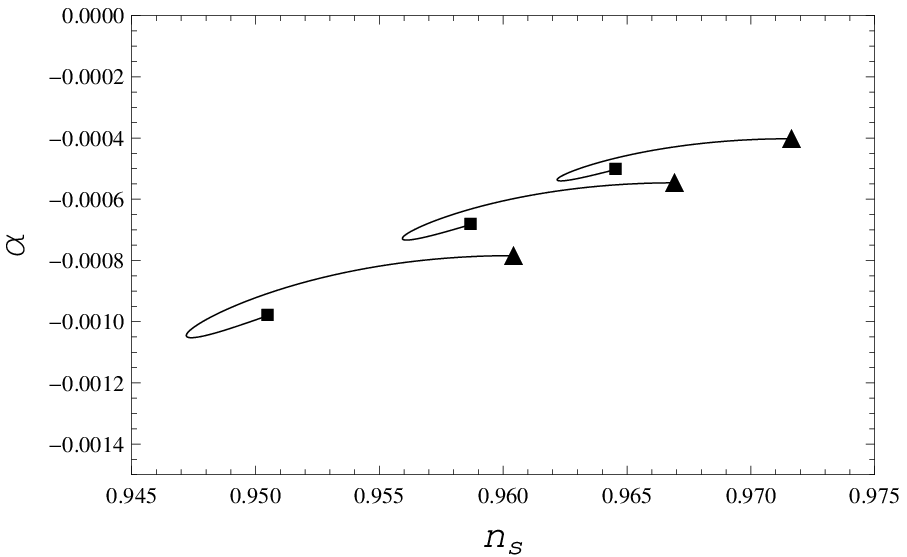}
\end{center}
\caption{
The inflationary predictions for the quadratic potential model: $n_s$ vs. $r$ (left panel) and $n_s$ vs. $\alpha$ (right panel)
  for various $M_5$ values with $N_0=50$, $60$ and $70$ (from left to right), along with the contours 
  (at the confidence levels of 68\% and 95\%)   
   given by the Planck 2015 results ({\it Planck} TT+lowP)~\cite{Planck2015}. 
The black triangles are the predictions of the textbook quadratic potential model in the standard cosmology, 
  which are reproduced for $M_5 \gtrsim 1$. 
As $M_5$ is lowered, the inflationary predictions approach the values in Eq.~(\ref{phi2_BClimit}), denoted by the black squares.   
In each line, the turning point appears for $V(\phi_0)/\rho_0 \simeq 1$. 
}
\label{fig:phi2}
\end{figure}

\begin{figure}[htbp]
\begin{center}
\includegraphics[width=0.45\textwidth,angle=0,scale=1.05]{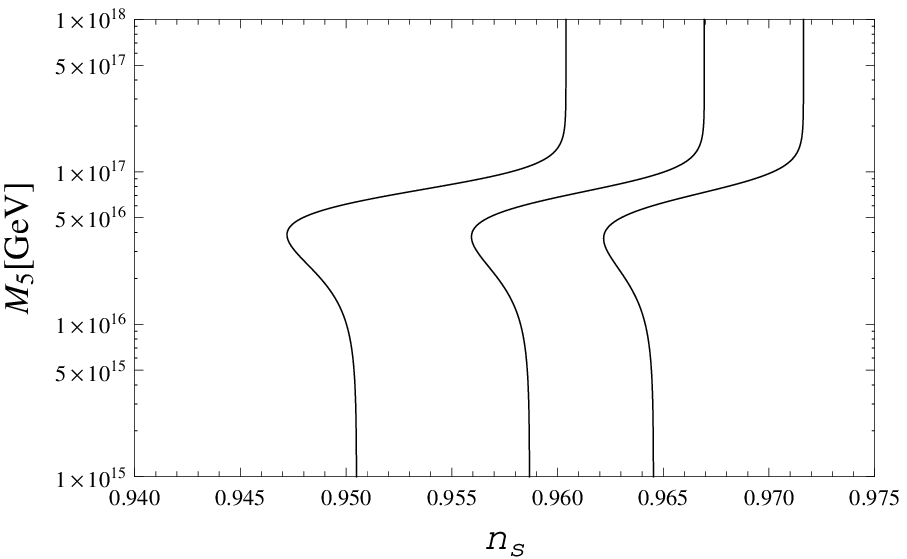} \hspace{0.5cm}
\includegraphics[width=0.45\textwidth,angle=0,scale=1.05]{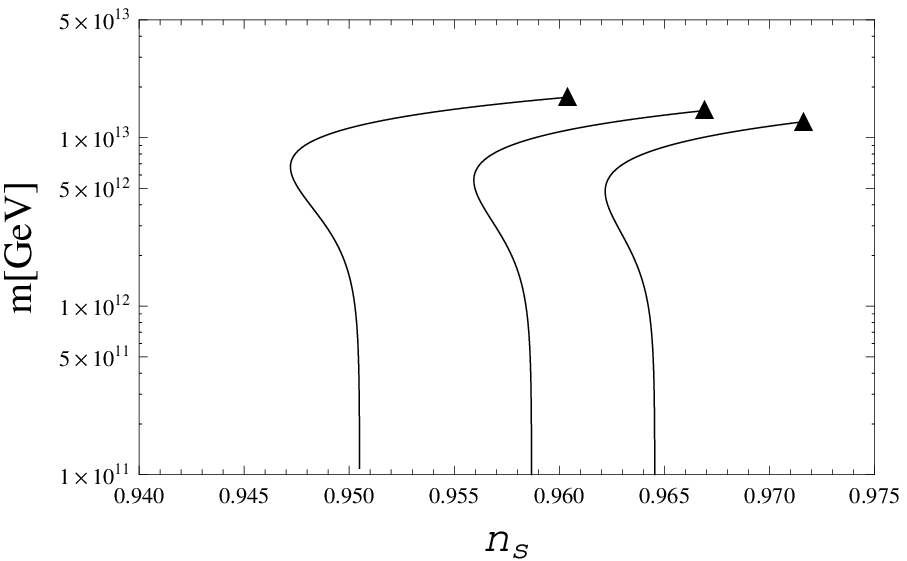}
\end{center}
\caption{
Relations between $n_s$ and $M_5$ (left panel) and between  $n_s$ and $m$ (right-panel), 
  for $N_0=50$, $60$ and $70$ from left to right. 
The black triangles denote the predictions in the standard cosmology.
}
\label{fig:phi2_mass}
\end{figure}

We calculate the inflationary predictions for various values of $M_5$ with fixed e-folding numbers, 
  and show the results in Fig.~\ref{fig:phi2}. 
In the left panel, the inflationary predictions for $N_0=50$, $60$ and $70$ from left to right are shown, 
 along with the contours (at the confidence levels of 68\% and 95\%) 
  given by the Planck 2015 results ({\it Planck} TT+lowP)~\cite{Planck2015}. 
The black triangles represent the predictions of the quadratic potential model in the standard cosmology, 
  while the black squares are the predictions in the limit of $V/\rho_0 \gg 1$. 
We see that the quadratic potential model is not favored by the Planck 2015 results, 
  and the brane-world cosmological effect makes the situation worse. 
For $N_0=70$, we have found an lower bound on $M_5[{\rm GeV}] \geq 9.98 \times 10^{16}$  
  for the inflationary prediction to stay inside of the Planck contour at 95\% C.L. 
The results for the running of the spectral index ($n_s$ vs. $\alpha$) is shown in the right panel, 
  for $N_0=50$, $60$, $70$ from left to right. 
The predicted $|\alpha|$ values are very small and consistent with the Planck 2015 results~\cite{Planck2015}, 
   $\alpha=-0.0126^{+0.0098}_{-0.0087}$ ({\it Planck} TT+lowP).   
We also show our results for the 5-dimensional Planck mass ($n_s$ vs. $M_5$) 
  and the inflaton mass ($n_s$ vs. $m$) in Fig.~\ref{fig:phi2_mass}. 
In each solid line, the turning point appears for $V(\phi_0)/\rho_0 \simeq 1$.

Next we analyze the textbook quartic potential model, 
\begin{eqnarray}
V =\frac{\lambda}{4!} \phi^4. 
\end{eqnarray}
In the standard cosmology, we find the following inflationary predictions:
\begin{eqnarray}
n_s=1-\frac{6}{2 N_0+3},  \; \; 
r=\frac{32}{2 N_0+3}, \; \; 
\alpha= - \frac{12}{(2 N_0+3)^2}. 
\end{eqnarray}
The quartic coupling ($\lambda$) is determined by the power spectrum measured by the Planck satellite experiment, 
 ${\cal P}_{\cal S}(k_0)=2.196 \times 10^{-9}$ at the pivot scale $k_0=0.002$ Mpc$^{-1}$,  as 
\begin{eqnarray}
  \lambda = 8.46 \times 10^{-13}  \left(  \frac{123}{2 N_0+3}\right)^3. 
\end{eqnarray}

When the limit $V/\rho_0 \gg 1$ is satisfied during the inflation, we find in the brane-world cosmology 
\begin{eqnarray}
n_s=1-\frac{9}{3 N_0+2},  \; \; 
r=\frac{48}{3 N_0+ 2}, \; \; 
\alpha= - \frac{27}{(3 N _0+2)^2}. 
\label{phi4_BClimit}
\end{eqnarray}
The predicted values are very close to those in the standard cosmology for $N_0 \gg 1$. 
Using ${\cal P}_{\cal S}(k_0)=2.196 \times  10^{-9}$, we find 
\begin{eqnarray}
 \lambda = 3.23 \times 10^{-14} \left(  \frac{182}{3 N_0+ 2}\right)^{3} ,  
\end{eqnarray} 
which is independent of $M_5$.

\begin{figure}[htbp]
\begin{center}
\includegraphics[width=0.45\textwidth,angle=0,scale=1]{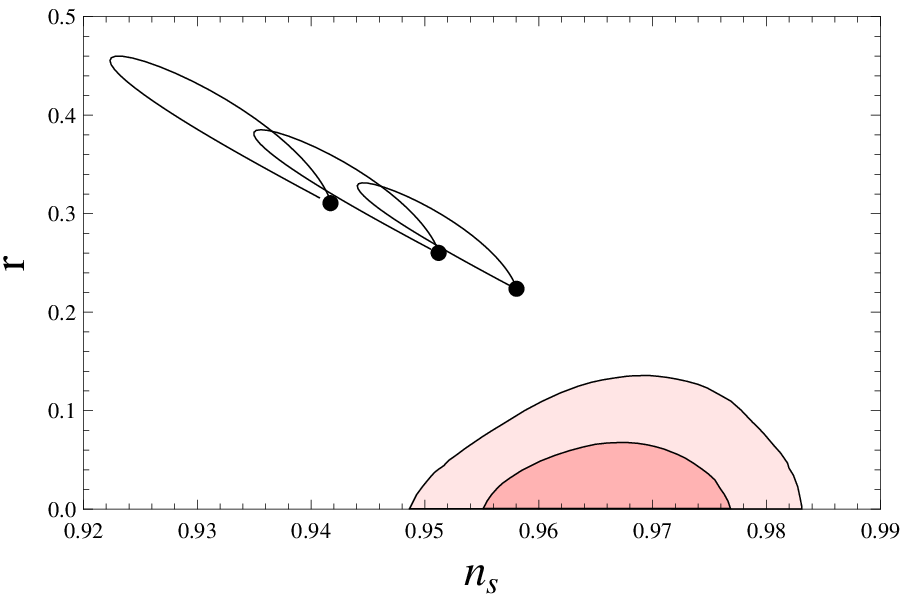} \hspace{0.5cm}
\includegraphics[width=0.45\textwidth,angle=0,scale=1.05]{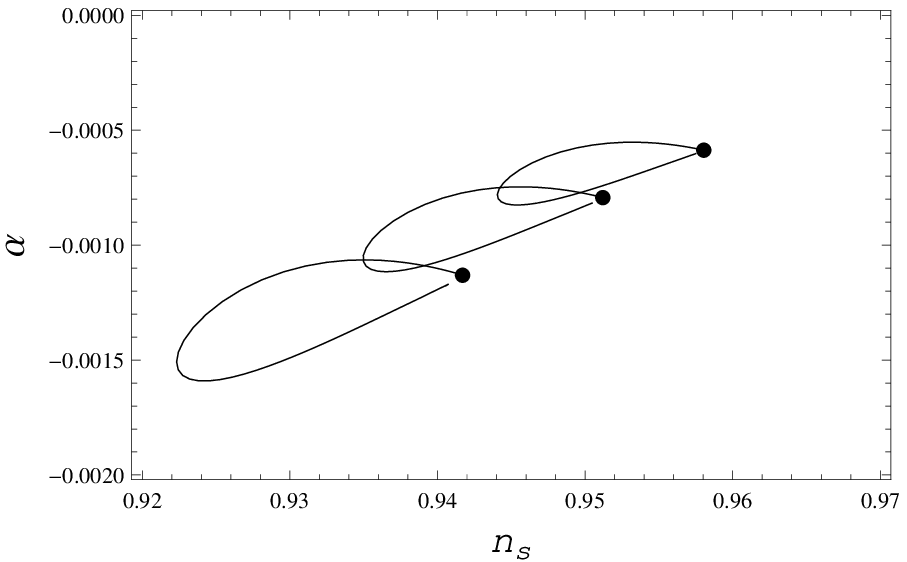}
\end{center}
\caption{
The inflationary predictions for the quartic potential model: $n_s$ vs. $r$ (left panel) and $n_s$ vs. $\alpha$ (right panel)
  with various $M_5$ values for $N_0=50$, $60$ and $70$ (from left to right), along with the contours 
  (at the confidence levels of 68\% and 95\%) 
   given by the Planck 2015 results ({\it Planck} TT+lowP)~\cite{Planck2015}. 
The black points are the predictions of the textbook quartic potential model in the standard cosmology, 
  which are reproduced for $M_5 \gtrsim 1$. 
As $M_5$ is lowered, the inflationary predictions approach the values in Eq.~(\ref{phi4_BClimit}) 
 which are in fact very close to those in the standard cosmology.  
In each line, the turning point appears for $V(\phi_0)/\rho_0 \simeq 1$. 
}
\label{fig:phi4}
\end{figure}

We calculate the inflationary predictions for various values of $M_5$ with fixed e-folding numbers. 
Our results are shown in Fig.~\ref{fig:phi4}. 
In the left panel, the inflationary predictions for $N_0=50$, $60$, $70$ from left to right are shown, 
 along with the contours (at the confidence levels of 68\% and 95\%) 
 given by the Planck 2015 results ({\it Planck} TT+lowP)~\cite{Planck2015}. 
The results for the running of the spectral index ($n_s$ vs. $\alpha$) is shown in the right panel, 
  for $N_0=$50, 60, 70 from left to right. 
The black points represent the predictions in the standard cosmology.  
In Fig.~\ref{fig:phi4}, the inflationary predictions are moving anti-clockwise along the contours as $M_5$ is lowered. 
The turning point on each contour appears for $V(\phi_0)/\rho_0 \simeq 1$, and as $M_5$ is further lowered, 
 the inflationary predictions go back closer to those in the standard cosmology. 
The quartic potential model is clearly disfavored by the Planck 2015 results, 
  and the brane-world cosmological effect cannot improve the fit.

\section{Higgs potential model}
Next we consider an inflationary scenario based on the Higgs potential of the form~\cite{Higgs_potential_model}
\begin{eqnarray}
V= \frac{\lambda}{8} \left( \phi^2 -v^2 \right)^2, 
\end{eqnarray}
where $\lambda$ is a real, positive coupling constant, $v$ is the VEV of the inflaton $\phi$. 
For simplicity, we assume the inflaton is a real scalar in this paper. 
It is easy to extended this model to the Higgs model 
  in which the inflaton field breaks a gauge symmetry by its VEV. 
See, for example, Refs.~\cite{hp_corrections, BL_inflation} for recent discussion, 
  where quantum corrections of the Higgs potential are also taken into account.

For analysis of this inflationary scenario, we can consider two cases for the initial inflaton VEVs:
  (i) $\phi_0 < v$ and (ii) $\phi_0 > v$. 
In the case (i), the inflationary prediction for the tensor-to-scalar ratio is found to be small for $v < 1$, 
  since the potential energy during inflation never exceeds $\lambda v^4/8$. 
This Higgs potential model reduces to the textbook quadratic potential model in the limit $v \gg 1$.  
To see this, we rewrite the potential in terms of $\phi=\chi +v$ with an inflaton $\chi$ around the potential minimum at $\phi=v$, 
\begin{eqnarray}
V= \frac{\lambda}{8} \left( 4 v^2 \chi^2 +4 v \chi^3 +\chi^4 \right). 
\end{eqnarray}
It is clear that if a condition, $\chi_0/v \ll 1$, for an initial value of inflaton ($\chi_0$) is satisfied, 
  the inflaton potential is dominated by the quadratic term. 
Recall that $\chi_0 > 1$ in the textbook quadratic potential model and hence $v \gg 1$ is necessary to satisfy the condition. 
In the case (ii), it is clear that the model reduces to the textbook quartic potential model for $ v \ll 1$. 
For the limit $v \gg 1$, we apply the same discussion in the case (i), so that the model reduces to the textbook quadratic potential model. 
Therefore, the inflationary predictions of the model in the case (ii) interpolate the inflationary predictions 
  of the textbook quadratic and quartic potential models by varying the inflaton VEV from $v=0$ to $v \gg 1$.

We now consider the brane-world cosmological effects on the Higgs potential model. 
As we have observed in the previous section, the inflationary predictions of the textbook models 
  are dramatically altered in the brane-world cosmology. 
The quadratic potential model is marginally consistent with the Planck 2015 results, 
  while the quartic potential model is disfavored.   
Thus, in the following, we concentrate on the case (i) with $\phi_0 < v$. 
Even in the brane-world cosmology, the above discussion for (ii) is applicable, namely, 
  the Higgs potential model reduces to  the textbook models for the very small or large VEV limit 
  and  the inflationary predictions for (ii) interpolate those in the two limiting cases. 
Thus, once we have obtained the inflationary predictions for the case (i),  we can imaginary interpolate them 
  to the results for the quartic potential model presented in the previous section.

\begin{figure}[htbp]
\begin{center}
\includegraphics[width=0.45\textwidth,angle=0,scale=1.0]{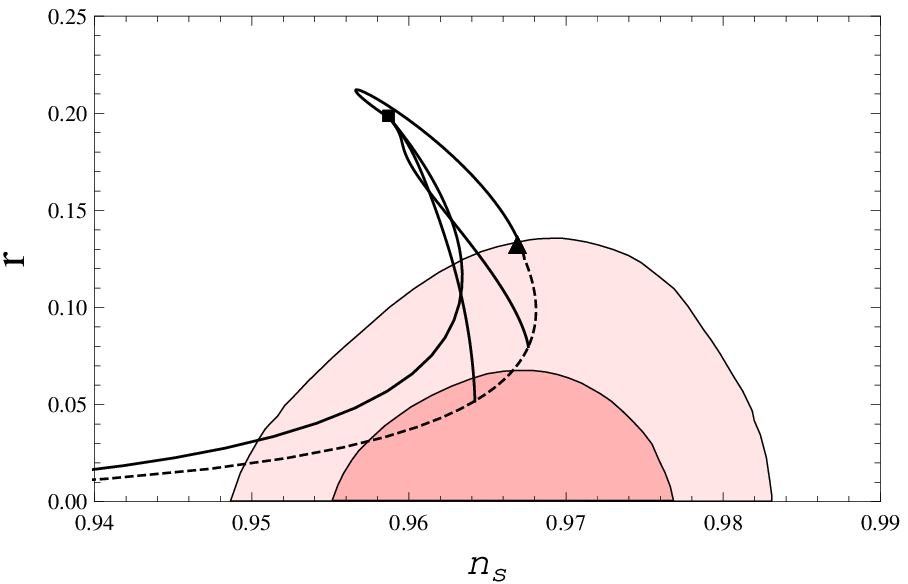} \hspace{0.5cm}
\includegraphics[width=0.45\textwidth,angle=0,scale=1.05]{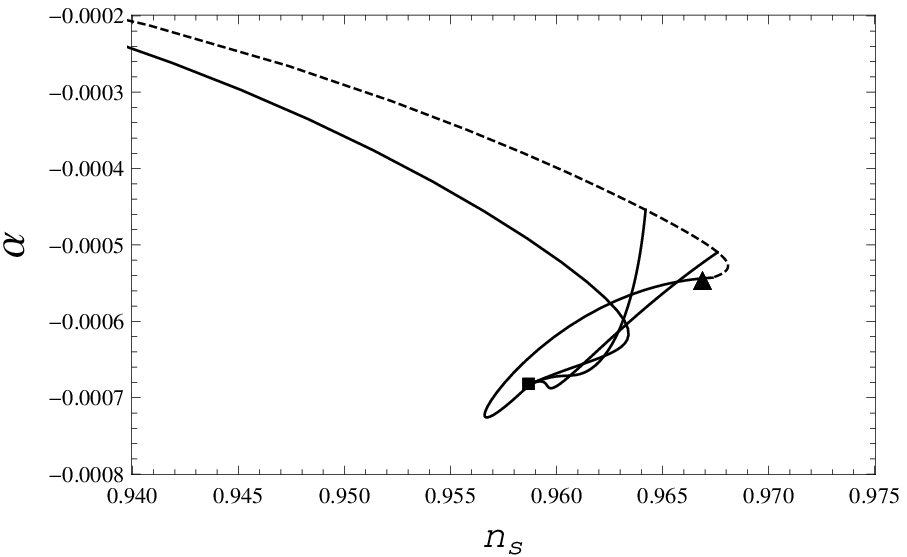}
\end{center}
\caption{
The inflationary predictions for the Higgs potential model: $n_s$ vs. $r$ (left panel) and $n_s$ vs. $\alpha$ (right panel)
  for various $M_5$ values with fixed $v=10$, $20$, $30$ and $200$ from left to right, along with the contours 
  (at the confidence levels of 68\% and 95\%) given by the Planck 2015 results ({\it Planck} TT+lowP).  
Here we have fixed the number of e-foldings $N_0=60$. 
The dashed line denotes the inflationary predictions for various values of $v$ from $10$ to $200$ in the standard cosmology.
As $v$ is raised, the predicted values of $n_s$ and $r$  approach those of the quadratic potential model along the dashed line 
  (the position marked by the the black triangle).
For $M_5 \gtrsim 1$, the brane-world cosmological effects are negligible, and the predicted values of $n_s$ and $r$ 
  lie on the dashed line. 
As $M_5$ is lowered, the inflationary predictions are deviating from the values on the dashed line, 
   and all solid lines are converging to the point marked by the black squares, which are the predictions 
   of the quadratic potential model for $M_5 \ll 1$. 
}
\label{fig:hp}
\end{figure}

\begin{figure}[htbp]
\begin{center}
\includegraphics[width=0.45\textwidth,angle=0,scale=1.02]{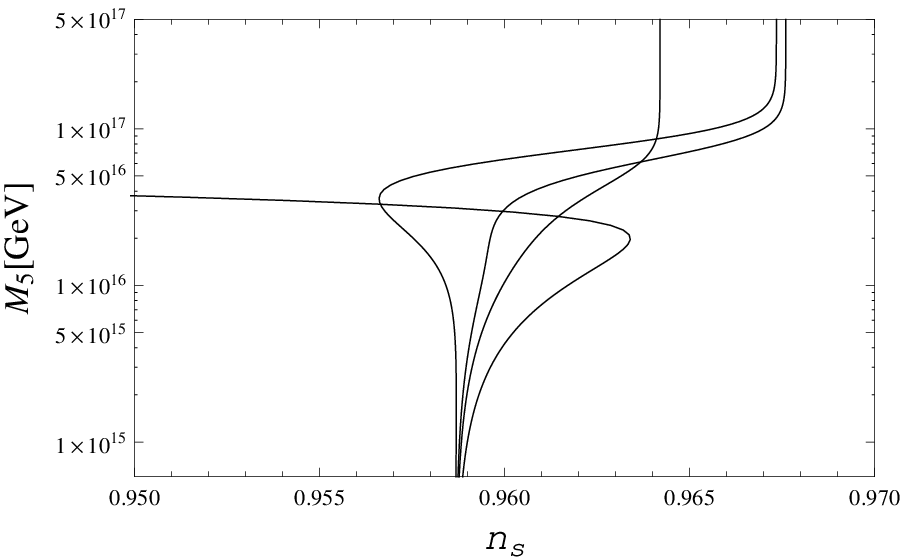} \hspace{0.5cm}
\includegraphics[width=0.45\textwidth,angle=0,scale=1.02]{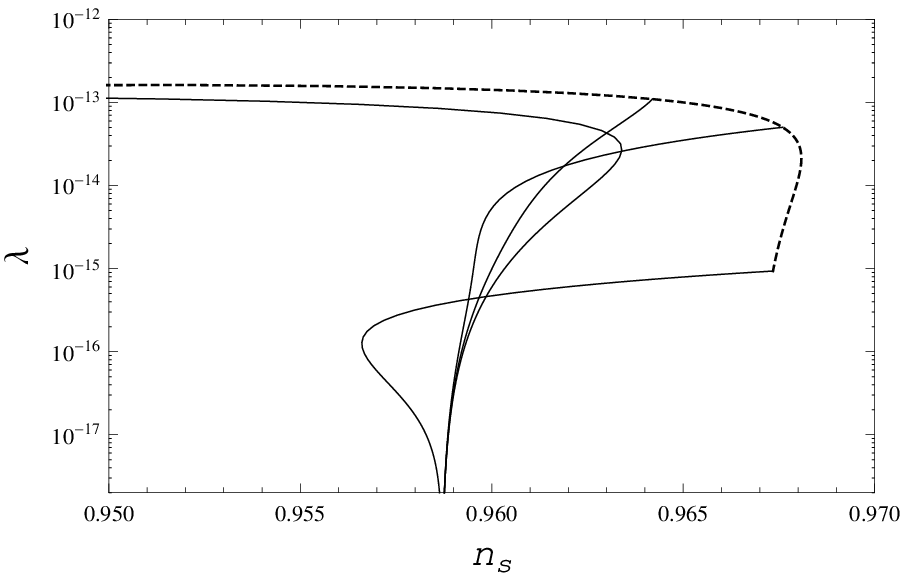}
\end{center}
\caption{
Relations between $n_s$ and $M_5$ (left panel) and between $n_s$ and $\lambda$ (right panel), 
  corresponding to Fig.~\ref{fig:hp}. 
}
\label{fig:hp_mass}
\end{figure}

We calculate the inflationary predictions for various values of $v$ and $M_5$, 
   and the results are shown in Fig.~\ref{fig:hp} for $N_0=60$. 
The dashed line denotes the results in the standard cosmology for various values of $v$. 
The black triangles represent the inflationary predictions in the quadratic potential model, 
  and we have confirmed that the inflationary predictions are approaching the triangles along the dashed line, as $v$ is raised. 
The solid lines show the results for various values of $M_5$ with $v=10$, $20$, $30$ and $200$ from left to right. 
For $M_5 \gtrsim 1$, the brane-world effect is negligible and the inflationary predictions lie on the dashed line. 
As $M_5$ is lowered, the inflationary predictions are deviating from those in the standard cosmology 
  and they approach the values obtained by the quadratic potential model in the brane-world cosmology 
  (shown as the black squares). 
We can see that the brane-world cosmological effect enhances the tensor-to-scalar ratio.  
For $v=20$, the inflationary prediction of the standard cosmology is outside of the contour, 
  but the brane-world cosmological effect alters the predictions to become consistent 
  with the Planck results for $2.07 \times 10^{16} \leq M_5[{\rm GeV}] \leq 3.70 \times 10^{16}$.  
For $v=20$, $30$ and $200$, the brane-world effect pushes the inflationary predictions 
  outside of the contours, so that there are lower bounds on $M_5$. 
We have found $M_5[{\rm GeV}]/10^{16} \geq 4.58$, $6.68$ and $13.5$ for $v=20$, $30$ and $200$, respectively, 
  in order to keep the inflationary reductions inside of the Planck contour at 95\% C.L. 
Fig.~\ref{fig:hp_mass} shows corresponding results in  ($n_s, M_5$)-plane (left panel) 
  and ($n_s, \lambda$)-plane (right panel). 
In the right panel, the dashed line denotes the results in the standard cosmology.

\section{Coleman-Weinberg potential}
Finally, we discuss an inflationary scenario based on a potential with a radiative symmetry breaking~\cite{Shafi_Vilenkin} 
  via the Coleman-Weinberg mechanism~\cite{Coleman_Weinberg}. 
We express the Coleman-Weinberg potential of the form, 
\begin{eqnarray}
 V= \lambda \phi^4 \left[ \ln \left( \frac{\phi}{v}\right)-\frac{1}{4}\right]+\frac{\lambda v^4}{4}, 
\end{eqnarray}
where $\lambda$ is a coupling constant, and $ v$ is the inflaton VEV. 
This potential has a minimum at $\phi=v$ with a vanishing cosmological constant. 
The inflationary predictions of the Coleman-Weinberg potential model 
 in the brane-world cosmology have recently been analyzed in \cite{CW_BC}. 
However, the modification of the power spectrum of tensor perturbation in Eq.~(\ref{PT}), 
 namely, the function $F$ was not taken into account in the analysis. 
In the following we will correct the results in \cite{CW_BC} by taking the function $F$ into account.  
We will see an enhancement of the tensor-to-scalar ratio due to the function $F$.

Analysis for the Coleman-Weinberg potential model is analogous to the one of the Higgs potential model 
   presented in the previous section. 
As same in the Higgs potential model, we can consider two cases, (i) $\phi_0 < v$ and (ii) $\phi_0 >v$, 
   for the initial VEV of the inflaton. 
In the same reason as in our discussion about the Higgs potential model, we only consider the case (i) 
  also for the Coleman-Weinberg potential model. 
Since the inflationary predictions in the case (ii) interpolate the predictions of the quartic potential model 
  to those of the quadratic potential model as $v$ is raised (with a fixed $M_5$), 
  one can easily imagine the results in the case (ii).

\begin{figure}[htbp]
\begin{center}
\includegraphics[width=0.45\textwidth,angle=0,scale=1.0]{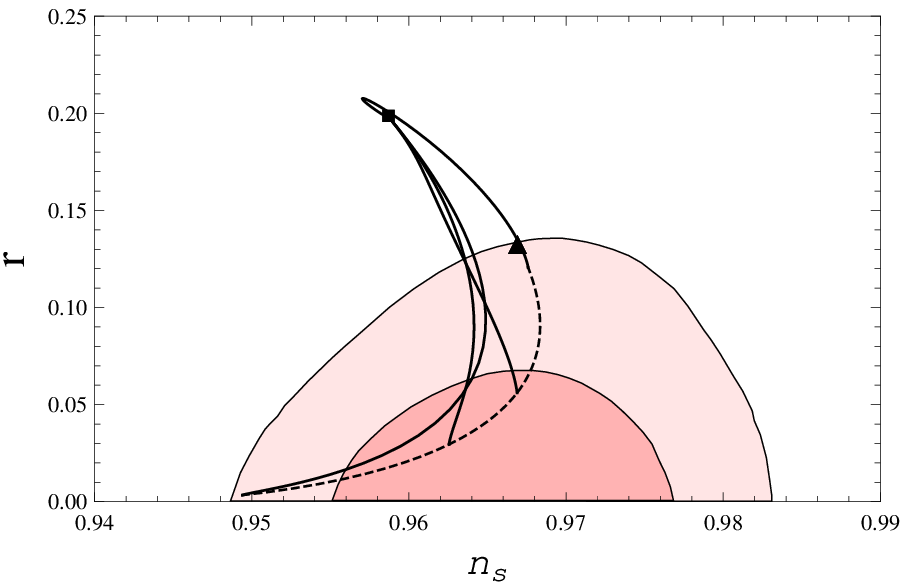} \hspace{0.5cm}
\includegraphics[width=0.45\textwidth,angle=0,scale=1.05]{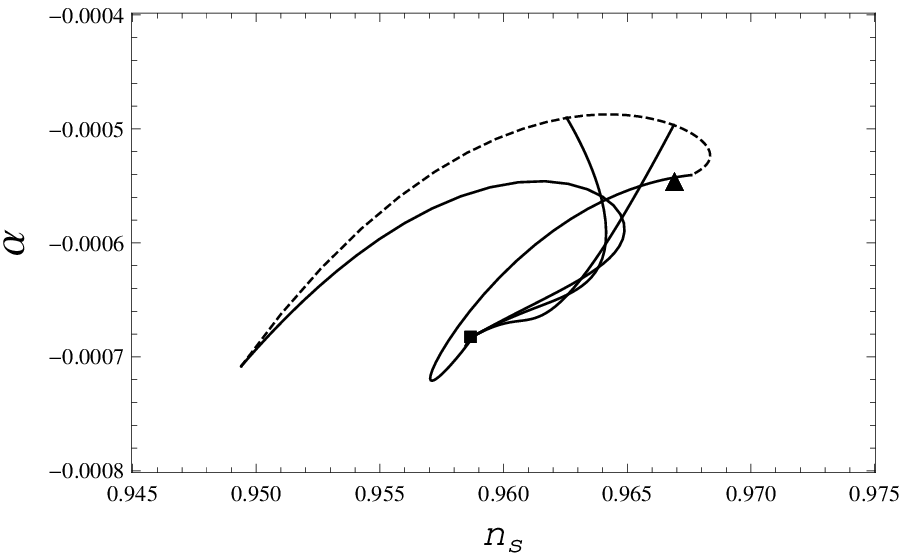}
\end{center}
\caption{
Same as Fig.~\ref{fig:hp} but for the Coleman-Weinberg potential model. 
}
\label{fig:CW}
\end{figure}

\begin{figure}[htbp]
\begin{center}
\includegraphics[width=0.45\textwidth,angle=0,scale=1.02]{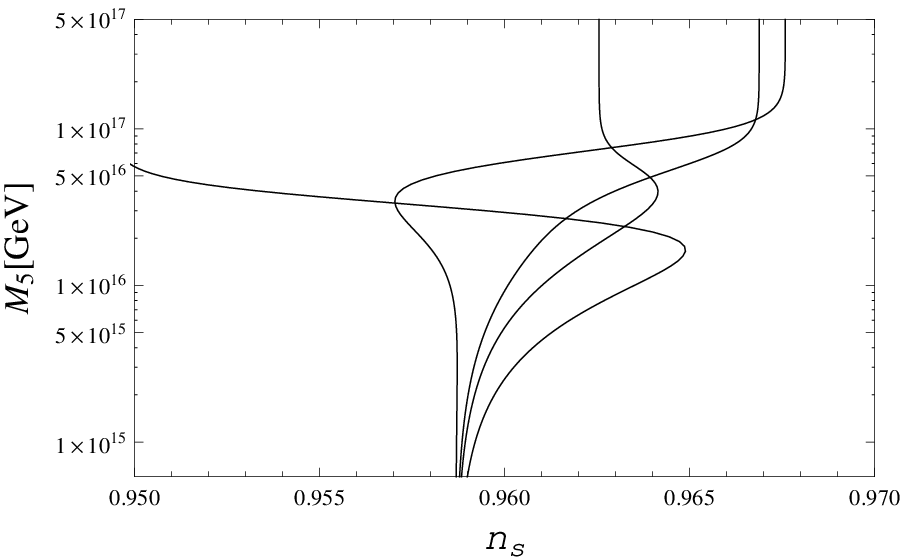} \hspace{0.5cm}
\includegraphics[width=0.45\textwidth,angle=0,scale=1.02]{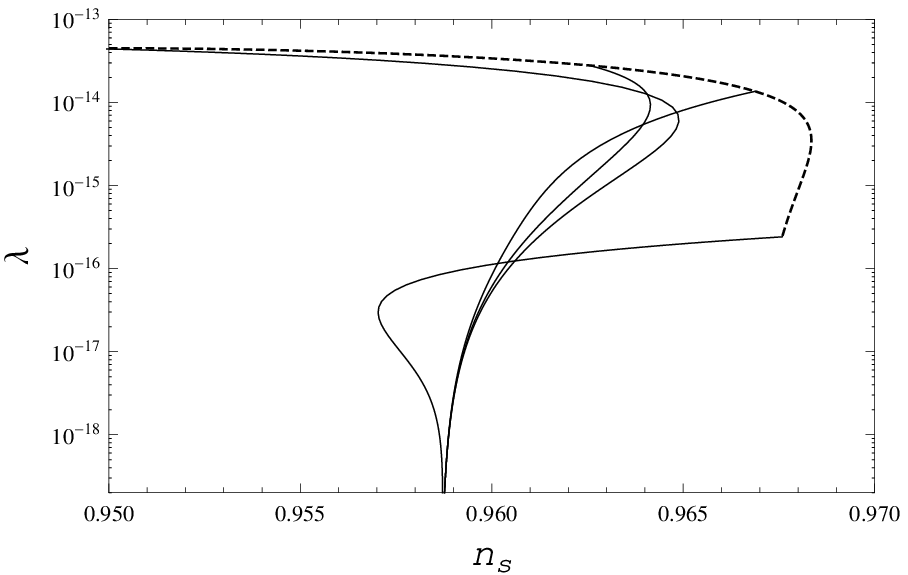}
\end{center}
\caption{
Same as Fig.~\ref{fig:hp_mass} but for the Coleman-Weinberg potential model. 
}
\label{fig:CW_mass}
\end{figure}

We show in Fig.~\ref{fig:CW} the inflationary predictions of the Colman-Weinberg potential model for various values 
 of $v$ and $M_5$ with $N_0=60$.  
The dashed line denotes the results in the standard cosmology for various values of $v$. 
The black triangles denote the results in the quadratic potential model, and we have confirmed that 
  the dashed line approaches the triangles as $v$ is increasing.  
The solid lines show the results for various values of $M_5$ with $v=10$, $20$, $30$ and $200$ from left to right. 
For $M_5 \gtrsim 1$, the brane-world effect is negligible and the inflationary predictions lie on the dashed line. 
As $M_5$ is lowered, the inflationary predictions are deviating from those in the standard cosmology 
   to approach the values obtained by the quadratic potential model in the brane-world cosmology 
  (shown as the black squares). 
As in the Higgs potential model, for $M_5 \ll v$ the inflationary predictions of 
  the Coleman-Weinberg potential model approach those of the quadratic potential model 
  in the brane-world cosmology. 
We have found the lower bounds on $M_5$ for the inflationary predictions 
  to stay inside of the Planck contour at 95\% C.L. as 
  $M_5[{\rm GeV}]/10^{16} \geq 2.18$, $5.78$, $7.83$ and $11.7$ for $v=10$, $20$, $30$ and $200$, respectively.  
Fig.~\ref{fig:CW_mass} shows corresponding results in  ($n_s, M_5$)-plane (left panel) 
   and ($n_s, \lambda$)-plane (right panel). 
In the right panel, the dashed line denotes the results in the standard cosmology.

\section{Conclusions}
Motivated by the Planck 2015 results, 
  we have studied simple inflationary models based on the quadratic, quartic, Higgs and Coleman-Weinberg potentials 
  in the context of the brane-world cosmology. 
For the 5-dimensional Planck mass $M_5 < M_P$, the brane-world cosmological effect 
  alters inflationary predictions from those in the standard cosmology.  
In particular, the tensor-to-scalar ratio is enhanced. 
The textbook quadratic potential model is marginally consistent with the Planck 2015 results 
  at 95\% C.L., and the brane-world effect makes the fitting worse. 
The textbook quartic potential model is disfavored by the Planck results, and  
  the brane-world effect cannot improve the data fitting. 
The Higgs and Coleman-Weinberg potential models show interesting behavior. 
With a suitable choice of the inflaton VEV, we have inflationary predictions in the standard cosmology 
   which are consistent with the Planck 2015 results. 
The brane-world cosmological effect pushes the predictions away from the allowed regain, 
  so that we have found lower bounds on $M_5$ in order to keep the inflationary predictions 
  inside of the allowed region. 
On the other hand, inflationary predictions in the standard cosmology laying outside of the Planck allowed region 
  can be kicked in the allowed region by the brane-world cosmological effect 
  with a suitable choice of $M_5$.

\section*{Acknowledgments}
We would like to thank Andy Okada for his encouragements.  
The work of N.O. is supported in part by the United States Department of Energy. 


\end{document}